\documentclass[11pt]{article}
\usepackage{graphicx,graphics,latexsym,verbatim,epsfig,subfigure}
\usepackage{mathrsfs}
\usepackage{amssymb,amsmath,amsthm}
\usepackage{algpseudocode}
\usepackage{algorithmicx}
\usepackage{algorithm}
\usepackage{url}
\newtheorem{lemma}{Lemma}

\newtheorem{example}{Example}

\newtheorem{corollary}{Corollary}

\newtheorem{proposition}{Proposition}

\begin{document}
\title{Simulation-based reachability analysis for nonlinear systems using componentwise contraction properties}
\author{Murat Arcak and John Maidens \\ Department of Electrical Engineering and Computer Sciences\\ University of California, Berkeley}
\maketitle

\begin{abstract}
A shortcoming of existing reachability approaches for nonlinear
systems is the poor scalability with the number of continuous state
variables. To mitigate this problem
we present a simulation-based approach where we first sample a number
of trajectories of the system and next establish bounds on the
convergence or divergence between the samples and neighboring
trajectories. We compute these bounds using contraction theory and
reduce the conservatism by partitioning the state vector into several
components and analyzing contraction properties separately in each
direction. Among other benefits this allows us to analyze the effect of constant but uncertain
parameters by treating them as state variables and partitioning them into
a separate direction.  We next present a numerical procedure to search for weighted norms that yield a prescribed contraction rate, which can be incorporated in the reachability algorithm to adjust the weights to minimize the growth of the reachable set.
\end{abstract}

\section{Introduction}
Reachability analysis is critical for testing and verification of control systems \cite{Kap17}, and for formal methods-based control synthesis
where
reachability dictates the transitions in a
discrete-state abstraction of a system with continuous dynamics \cite{tabuada}.
 Existing reachability approaches for nonlinear systems include level set methods \cite{Mitchell05},  linear or piecewise linear approximations of nonlinear models followed by linear reachability techniques  \cite{Althoff08, Chutinan03},  interval Taylor series methods \cite{Lin07, Neher07}, and differential inequality methods \cite{Lakshmikantham69, Scott13}. 
 However, these results typically scale poorly with the number of continuous state variables, limiting their applicability in practice. 

On the other hand trajectory-based approaches \cite{Donze07, Huang12, Julius09} scale well with the state dimension, as they take advantage of inexpensive numerical simulations and are naturally parallelizable. In \cite{MaiArc15} we leveraged concepts from {\it contraction theory} \cite{LS98, Sontag10} to develop a
new trajectory-based approach where we first sample a
number of trajectories of the system and next establish bounds on
the divergence between the samples and neighboring trajectories. We then use these bounds to provide a guaranteed over-approximation of the reachable set.
Unlike \cite{Huang12} that uses Lipschitz constants to bound the divergence between trajectories we use matrix measures that can take negative values, thus allowing for convergence of trajectories and reducing the conservatism in the over-approximation.  Another related reference,
\cite{Donze07}, uses sensitivity equations to
track the convergence or divergence properties along simulated trajectories; however, this approach
does not guarantee that the computed approximation contains the true reachable set.

In this note we generalize \cite{MaiArc15} by partitioning the state vector into several components and analyzing the growth or contraction properties in the direction defined by each component.  Unlike \cite{MaiArc15} which searches for
a single growth or contraction rate to cover every direction of the state space, the new approach takes advantage of directions that offer more favorable rates.  
With this generalization we can now analyze the effect of constant but uncertain parameters by treating them as state variables and 
partitioning them into a separate direction along which no growth occurs.
A related approach is employed in \cite{SCOTS} where every state variable defines a separate direction; however, this may
 lead to overly conservative results since the dynamics associated with multiple state variables may possess a more favorable rate than the individual state variables in isolation.  

In Section \ref{sec:contraction} we present the main contraction result and a corollary that serves as the starting point for the reachability algorithm.
In Section \ref{sec:algorithm} we detail the algorithm and demonstrate with an example that it can significantly reduce the conservatism in 
\cite{MaiArc15} and \cite{SCOTS}.  In Section \ref{sec:weights}  we present a numerical procedure to search for weighted norms that yield a prescribed contraction rate, which can be incorporated in the reachability algorithm to adjust the weights to minimize the growth of the reachable set as it propagates through time.

In the sequel we make use of matrix measures, as defined in \cite{desoer}. Let $| \cdot |$ be a norm on $\mathbb{R}^n$ and let $\| \cdot \|$ denote the induced matrix norm.  The measure $\mu(A)$ of a matrix $A \in \mathbb{R}^{n \times n}$ is the upper right-hand derivative of $\| \cdot \|$ at $I \in \mathbb{R}^{n \times n}$ in the direction of $A$: 
\begin{equation}\label{measure}
\mu(A)\triangleq \lim_{h\rightarrow 0^+}\frac{\| I+hA\|-1}{h}.
\end{equation}
Unlike a norm the matrix measure can take negative values, as evident in the table below.
\begin{table}[h!]
  \centering
  \resizebox{.8\columnwidth}{!}{%
  \begin{tabular}{|c|c|c|}
  \hline
  Vector norm & Induced matrix norm & Induced matrix measure \\
  \hline
  $|x|_1 = \sum_j |x_j| $ & $\|A\|_1=  \max_j \sum_i |a_{ij}|$ & $\mu_1(A) = \max_j \Big( a_{jj} + \sum_{i \not= j} |a_{ij}| \Big)$ \\
  $|x|_2 = \sqrt{\sum_j x_j^2}$ & $\|A\|_2  = \sqrt{ \max_j \lambda_j(A^TA) }$  & $\mu_2(A) = \max_j \frac{1}{2} \Big( \lambda_j (A + A^T) \Big)$  \\
  $|x|_\infty = \max_j |x_j|$ & $\|A\|_\infty =  \max_i \sum_j |a_{ij} |$ & $\mu_\infty(A) = \max_i\Big( a_{ii} + \sum_{j \not= i} |a_{ij}| \Big)$ \\
  \hline
  \end{tabular}
  }
  \caption{\small Commonly used vector norms and their corresponding matrix norms and measures.}
  \label{tab:measures}
\end{table}

\section{Componentwise Contraction}\label{sec:contraction}
Consider the nonlinear dynamical system
\begin{equation}\label{NL}
\dot{x}(t)=f(t,x(t)), \quad x(t)\in \mathbb{R}^n,
\end{equation}
where $f:[0,\infty) \times \mathbb{R}^n \mapsto \mathbb{R}^n$ is continuous in $t$ and continuously differentiable in $x$.  
We partition the state vector $x$ into $k$ components, $x=[x_1^T \cdots x_k^T]^T$, where $x_i\in \mathbb{R}^{n_i}$, $i=1,\dots,k$, and $n_1+\cdots+n_k=n$.
Likewise we decompose the $n\times n$ Jacobian matrix $J(t,x)=(\partial f/\partial x)(t,x)$  into conformal blocks $$J_{ij}(t,x)\in \mathbb{R}^{n_i\times n_j}, \quad i,j=1,\dots,k.$$

The following proposition gives a growth bound between two trajectories of the system (\ref{NL}).
Variants of this proposition appear in \cite{KapZgl09,RusDibSon13,ReiWebRun17}; we provide an independent proof in Section \ref{proof}.

\begin{proposition}\label{main}
Let $C\in \mathbb{R}^{k\times k}$ be a constant matrix such that 
\begin{equation}\label{bound}
C_{ij} \ge \left\{ \begin{array}{ll} \mu(J_{ii}(t,x)) & i=j \\ \|J_{ij}(t,x)\| & i\neq j \end{array} \right.
\end{equation}
for all $(t,x)\in [0,T]\times \mathcal{D}$ on some domain
$\mathcal{D}\subset \mathbb{R}^n$ .  
If $x(\cdot)$ and $z(\cdot)$ are two trajectories of (\ref{NL}) such that every trajectory 
starting on the line segment $\{s x(0)+(1-s)z(0): s \in [0,1]\}$ remains in $\mathcal{D}$ until time $T$, then
\begin{equation}\label{contract}
\left[\begin{array}{c} |x_1(t)-z_1(t)| \\ \vdots \\ |x_k(t)-z_k(t)| \end{array} \right] \le \exp(Ct) \left[\begin{array}{c} |x_1(0)-z_1(0)| \\ \vdots \\ |x_k(0)-z_k(0)| \end{array} \right] \quad \forall t\in [0,T],
\end{equation}
where $\le$ denotes element-wise inequality.\hfill $\Box$
\end{proposition}

We can use a different vector norm for each component in (\ref{contract}), say $|\cdot |_{p_i}$ for $x_i(t)-z_i(t)$, provided that we interpret (\ref{bound}) as
\begin{equation}\label{bound+}
C_{ij} \ge \left\{ \begin{array}{ll} \mu_{p_i}(J_{ii}(t,x)) & i=j \\ \|J_{ij}(t,x)\|_{p_i,p_j} & i\neq j, \end{array} \right.
\end{equation}
where $\mu_{p_i}(\cdot)$ is the matrix measure for $| \cdot |_{p_i}$, and $ \|\cdot \|_{p_i,p_j}$   is the mixed norm defined as
$$
\|A\|_{p_i,p_j}=\max_{|x|_{p_j}=1}|Ax|_{p_i}.
$$

We next derive a corollary to Proposition \ref{main} that is useful for reachability analysis.  
Let $\xi(t,x_0)$ denote the solution of (\ref{NL}) starting from $x_0$ at $t=0$, and define the reachable set at time $t$ from initial set $Z$ as
$$
Reach_t(Z)\triangleq \{\xi(t,z): z\in Z\}.
$$
Likewise define the reachable set over the time interval $[0,T]$ as
$$
Reach_{[0,T]}(Z)\triangleq \cup_{t \in [0,T]}Reach_t(Z).
$$
\begin{corollary}\label{prealgo}
Let $x(\cdot)$ be a trajectory of (\ref{NL}) and  define the norm ball of initial conditions 
$$
\mathcal{B}_{(\epsilon_1,\cdots,\epsilon_k)}(x(0))\triangleq \{ z: |x_1(0)-z_1|\le \epsilon_1, \cdots,  |x_k(0)-z_k|\le \epsilon_k\},
$$
centered at $x(0)$.
Suppose a coarse over-approximating set $\mathcal{D}\subset \mathbb{R}^n$ is available such that
\begin{equation}\label{coarse}
Reach_{[0,T]}(\mathcal{B}_{(\epsilon_1,\cdots,\epsilon_k)}(x(0))) \subset \mathcal{D}
\end{equation}
and $C\in \mathbb{R}^{k\times k}$ satisfies (\ref{bound}) for all $(t,x)\in [0,T]\times \mathcal{D}$.  
Then, 
\begin{equation}\label{ballbound}
Reach_{T}(\mathcal{B}_{(\epsilon_1,\cdots,\epsilon_k)}(x(0))) \subset \mathcal{B}_{(\delta_1,\cdots,\delta_k)}(x(T))
\end{equation}
where
\begin{equation}\label{ballexp}
\left[\begin{array}{c} \delta_1 \\ \vdots \\ \delta_k \end{array} \right] = \exp(CT) \left[\begin{array}{c} \epsilon_1\\ \vdots \\ \epsilon_k \end{array} \right].
\end{equation}
\mbox{} \hfill $\Box$
\end{corollary}

Corollary \ref{prealgo} relies on a coarse over-approximation $\mathcal{D}$ of the reachable set in (\ref{coarse}) to find a constant matrix $C$ satisfying (\ref{bound}).  
It then uses this $C$ in (\ref{ballbound})-(\ref{ballexp}) to find a more accurate over-approximation of the reachable set at the end of the time interval.
One can choose $\mathcal{D}$ to be a bounded invariant set for the system (\ref{NL}), or the entire state space 
if a global upper bound exists on the right-hand side of (\ref{bound}). For a less conservative estimate one can find a bound on each component of the vector field $f$ on an invariant set of interest,
\begin{equation}\label{fbound}
|f_i(t,x)|\le M_i, \quad i=1,\dots,k,
\end{equation}
and let
\begin{equation}\label{Dfunction}
\mathcal{D}=\mathcal{B}_{(\epsilon_1+M_1T,\cdots,\epsilon_k+M_kT)}(x(0)),
\end{equation}
which gives a tighter bound when the interval length $T$ is smaller.

\section{Simulation-based Reachability Algorithm}\label{sec:algorithm}
Given a sequence of simulation points $x[l]\triangleq x(t_l)$, $l=0,1,\dots,L$, Algorithm \ref{algo} below 
tracks the evolution of the initial norm ball along this
trajectory
by applying Corollary \ref{prealgo} along with the bound (\ref{Dfunction}) to each interval $[t_l,t_{l+1}]$, $l=0,1,\dots,L-1$.

\begin{algorithm}               
\footnotesize \caption{Algorithm for bounding reachable tube along a sample trajectory  \label{algo}      }          
\label{alg:local}                           
\begin{algorithmic}[1]                    
   \footnotesize \Require Vector $\epsilon=[\epsilon_1,\cdots, \epsilon_k]^T$ for the initial ball size, sequence of simulation points $x[l]\triangleq x(t_l)$, $l=0,1,\dots,L$, and bounds $M_1,\cdots, M_k$  as in (\ref{fbound})
    \State  Set $\delta[0] = \epsilon$ 
         \For{$l$ from 0 to $L-1$}
         \State Compute 
         matrix $C_l$ that satisfies (\ref{bound}) for \\ 
          \ \ \ \ \ \ \ \ \ \ $t_{l} \le t \le t_{l+1}$    and $x\in \mathcal{B}_{(\delta_1[l]+M_1(t_{l+1} - t_l),\cdots,\delta_k[l]+M_k(t_{l+1} - t_l))}(x[l])$. 
              \State Set $\delta[l+1] = \exp(C_l(t_{l+1} - t_l)) \delta[l]$
      \EndFor \\ 
\Return $\mathcal{B}_{\delta_1[l],\cdots,\delta_k[l]}(x[l])$, $l=1,\dots,L$
         \end{algorithmic}
\end{algorithm} 

A similar approach to reachability was pursued in \cite{MaiArc15}, using the special case of Proposition \ref{main} for $k=1$. The choice $k=1$ amounts to looking for a single growth or contraction rate to cover every direction of the state space, which is conservative when some directions offer more favorable rates than others.  The other extreme, $k=n$, used in \cite{SCOTS} can also lead to overly conservative results, since the dynamics associated with multiple state variables may possess a more favorable rate than the individual state variables in isolation.  The following example illustrates that intermediate choices of $k$ may give tighter bounds than the extremes $k=1$ and $k=n$.

\begin{example}\label{harmonic}
We consider the harmonic oscillator 
 \begin{eqnarray}\label{ho1}
\dot{p}(t)&=&\omega q(t) \\
\dot{q}(t)&=&-\omega p(t) \label{ho2}
\end{eqnarray}
and treat the constant frequency $\omega$ as a state variable satisfying
\begin{equation}\label{ho3}
\dot{\omega}(t)=0,
\end{equation}
so that we can view different values of $\omega$ as variations of the initial condition $\omega(0)$.  Thus, the state vector is $x=[p\  q\ \omega]^T$ and the  Jacobian matrix for (\ref{ho1})-(\ref{ho3}) is
\begin{equation}
\label{Jacob-ex}
J(x)=\begin{bmatrix} 0 & \omega & q \\-\omega & 0 & -p \\ 0 & 0 & 0 \end{bmatrix}.
\end{equation}
If we partition $x$ into $k=2$ components as
$x_1=[p \ q ]^T$ and $x_2=\omega$, then
\begin{equation}
J_{11}(x)=\begin{bmatrix} 0 & \omega \\-\omega & 0 \end{bmatrix} \quad J_{12}(x)=\begin{bmatrix} q \\-p \end{bmatrix} \quad  J_{21}(x)=\begin{bmatrix} 0 & 0 \end{bmatrix} \quad J_{22}(x)=0,
\end{equation}
and the matrix measures and norms induced by the Euclidean norm are
$$
\mu(J_{11}(x))=0, \quad \|J_{12}(x)\|=r\triangleq \sqrt{p^2+q^2}, \quad \|J_{21}(x)\|=0, \quad \mu(J_{22}(x))=0.
$$
Thus, 
$$
C=\begin{bmatrix} 0 & \bar{r} \\ 0 & 0\end{bmatrix}
$$
satisfies (\ref{bound}) on the invariant set $r\le \bar{r}$, and it follows from Corollary \ref{prealgo} 
that the
initial norm ball 
\begin{equation}\label{initial-ex}
\{ (p,q,\omega): (p-p(0))^2+(q-q(0))^2 \le \epsilon_1^2, \ |\omega-\omega(0)|\le \epsilon_2\}
\end{equation}
evolves to
\begin{equation}\label{final-ex}
\{ (p,q,\omega): (p-p(T))^2+(q-q(T))^2 \le \delta_1^2,\ |\omega-\omega(T)|\le \delta_2\},
\end{equation}
where $\omega(T)=\omega(0)$ is the nominal frequency with which the sample trajectory is obtained, and
\begin{equation}\label{final-ex+}
\delta_1=\epsilon_1+\bar{r}\epsilon_2T, \quad \delta_2=\epsilon_2
\end{equation}
from (\ref{ballexp}).  Note that (\ref{final-ex+}) correctly predicts the absence of growth in the $\omega$ direction, while accounting for the effect of frequency variation on $(p,q)$ by enlarging the radius of the corresponding ball to $\delta_1=\epsilon_1+\bar{r}\epsilon_2T$.
Algorithm \ref{algo} gives a tighter estimate of $\delta_1$ by applying Corollary \ref{prealgo}  along with the bound (\ref{Dfunction}) in every interval $[t_l,t_{l+1}]$, $l=0,1,\dots,L-1$, of the simulated trajectory.  A result of this algorithm is shown in Figure \ref{harmonicosc} (left) when $w=1$ and $\epsilon_2=0.02$, that is when a $\pm2\%$ uncertainty is allowed around the nominal frequency.  
The right panel shows the result with $\epsilon_2=0$, in which case there is no uncertainty in frequency and the radius of the norm ball remains constant.

In this example we applied Proposition \ref{main} by partitioning the state into $k=2$ components.
The alternative choice $k=1$ (no partition) amounts to searching for a single growth rate in each direction and fails to identify the lack of change in the $\omega$ direction. Indeed the matrix measure of (\ref{Jacob-ex}) is positive for any choice of the norm and, thus, the norm ball grows in every direction.  
The choice $k=3$ is also overly conservative because it misses the non-expansion property of the combined $(p,q)$ dynamics (\ref{ho1})-(\ref{ho2}), instead applying (\ref{ballexp}) with a matrix of the form
$$
C=\begin{bmatrix} 0 & \bar\omega &\bar{r} \\ \bar\omega & 0 & \bar{r} \\ 0 & 0 & 0\end{bmatrix},
$$
which falsely predicts a rapid growth of the norm ball in the $(p,q)$ direction even when no uncertainty is present in the frequency.

\begin{figure}[t]
\begin{center}
\includegraphics[width=.9\linewidth]{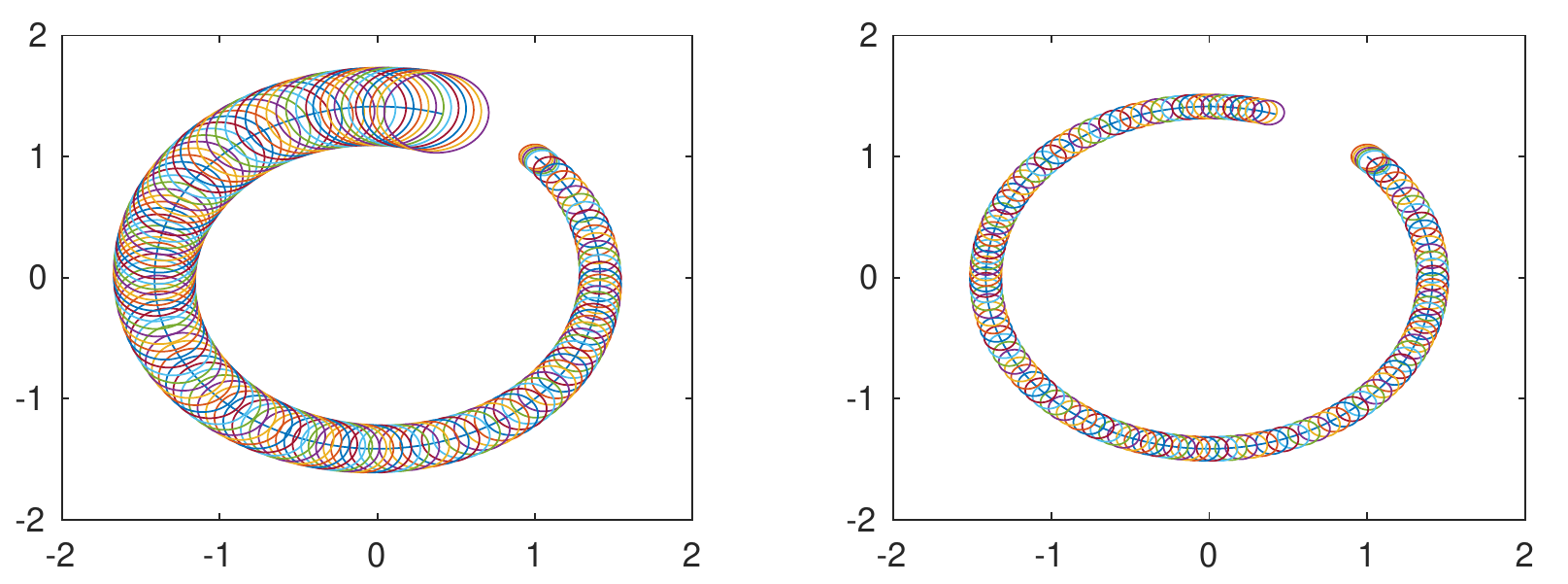}
\put(-100,-5){$p$}
\put(-325,-5){$p$}
\put(-430,80){$q$}
\put(-205,80){$q$}
\vspace{-.5cm}
\end{center}
\caption{\label{harmonicosc} Algorithm \ref{algo} applied to a sample trajectory of (\ref{ho1})-(\ref{ho3}) with $\omega=1$.  The initial norm ball is as in (\ref{initial-ex}) with $\epsilon_1=0.1$.  The plot on the left takes $\epsilon_2=0.02$, which means a $\pm2\%$ uncertainty around the nominal frequency $\omega=1$. 
This uncertainty leads to a growth of the radius of the norm ball in the $(p,q)$ direction.
The plot on the right takes $\epsilon_2=0$ and, since there is no uncertainty in the frequency, the radius of the norm ball in the $(p,q)$ direction remains constant.
The bound $M_1$ used in the algorithm is calculated on the invariant set $\sqrt{p^2+q^2}\le 2$.
}
\end{figure}
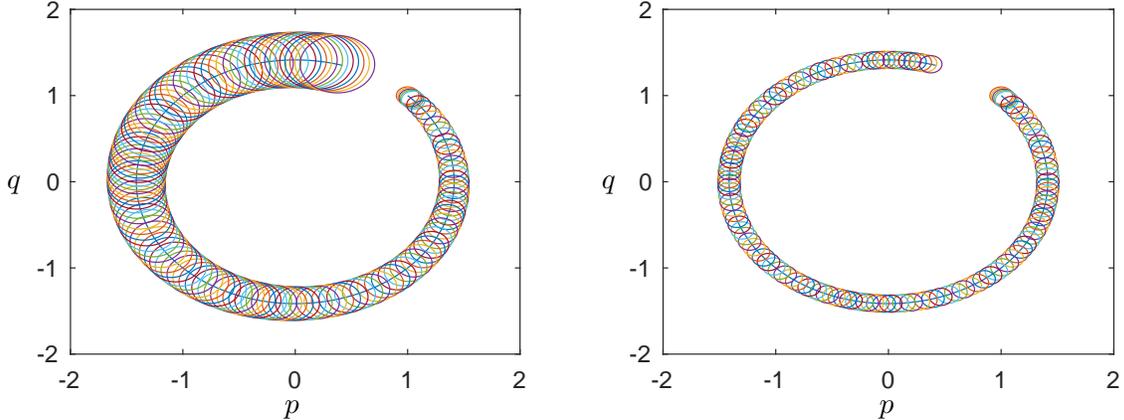

\end{example}

\section{Automatic Selection of Weighted Norms}\label{sec:weights}

In \cite{MaiArc15} we demonstrated that bounding the matrix measure induced by weighted 1-, 2- or $\infty$-norms can be expressed as constraints that are convex functions of the weights. We used this fact to develop a heuristic for minimizing growth of the reachable set as it propagates through time. The authors of \cite{Fan16} use the linear matrix inequality (LMI) constraint corresponding to the weighted 2-norm together with interval bounds on the Jacobian to argue that optimal bounds on the Jacobian matrix measure can be computed automatically by solving a sequence of semidefinite programs (SDPs). 

In this section, we demonstrate how these types of methods can be extended to the case of componentwise contraction. In Proposition \ref{prop:infinite_SDP} we show that for a fixed matrix $C$ the set of weight matrices for which a weighted 2-norm version of \eqref{bound} holds is a convex set that can be expressed as the conjunction of an infinite number of LMIs. Then in Proposition \ref{prop:finite_DSP} we show how polytopic bounds on the Jacobian can be exploited to compute an inner approximation of the set of feasible weight matrices, enabling weight matrices to be selected automatically using a standard numerical SDP solver. 

We begin with two Lemmas that demonstrate how weighted matrix measure and norm bounds can be expressed as LMIs. Throughout this section the inequality symbol $A \preceq B$ is used to denote that $B - A$ is a positive semidefinite matrix. 
\begin{lemma} (Lemma 2 of \cite{Aminzare13}) \label{measure_LMI}
If 
\[
\Gamma A + A^T \Gamma \preceq 2c \Gamma
\]
where $\Gamma$ is a positive definite matrix then $\mu(A) \le c$ in the norm $x \to |Px|_2$ where $P = \Gamma^{1/2} \succeq 0$. 
\end{lemma}

\begin{lemma} \label{norm_LMI}
If 
\[
A^T \Gamma_i A \preceq c^2 \Gamma_j 
\]
where $\Gamma_i$ and $\Gamma_j$ are positive definite matrices and $c \ge 0$ then $\|A\| \le c$ where $\| \cdot \|$ denotes the mixed norm
\[
\|A\| = \max_{ |P_j x|_2 = 1 } |P_i A x|_2
\]
where $P_i = \Gamma_i^{1/2}$ and $P_j = \Gamma_j^{1/2}$. 
\end{lemma} 
\begin{proof}
$A^T \Gamma_i A \preceq c^2 \Gamma_j$ implies that for all $x$ 
\[
|P_i Ax|_2^2 = x^T A^T \Gamma_i A x \le c^2 x^T \Gamma_j x = c^2 |P_j x|_2^2. 
\]
Thus 
\[
\|A\|^2 = \max_{|P_j x |_2 = 1} |P_i Ax|_2^2 \le c^2, 
\]
or equivalently $\|A\| \le c$.
\end{proof}

Combining Lemmas \ref{measure_LMI} and \ref{norm_LMI} along with the reasoning used to derive Equation \eqref{bound+}, we arrive at the following result: 
\begin{proposition} \label{prop:infinite_SDP}
Given a matrix $C \in \mathbb{R}^{k \times k}$, the search for weighted Euclidean norms $x_i \mapsto |P_i x_i|$ for $i = 1, \dots, k$ in which \eqref{bound} is satisfied can be formulated as a semidefinite program: 
\begin{equation}\label{infinite_SDP}
\begin{split}
 \textbf{find \quad} & \Gamma_1, \dots, \Gamma_k \\
 \textbf{subject to \quad} & \Gamma_i \succeq 0, 
 \quad \forall i = 1, \dots, k \\
 & \Gamma_i J_{ii}(t, x) + J_{ii}(t, x)^T \Gamma_i \preceq 2 c_{ii} \Gamma_i, 
 \quad \forall(t, x) \in [0, T] \times \mathcal{D}, 
 \quad \forall i = 1, \dots, k \\
 & J_{ij}(t, x)^T \Gamma_i J_{ij}(t, x) \preceq c_{ij}^2 \Gamma_j 
 \quad \forall(t, x) \in [0, T] \times \mathcal{D}, 
 \quad \forall j\neq i,\, i = 1, \dots, k
\end{split}
\end{equation}
where $\Gamma_i = P_i^T P_i$. 
\end{proposition}

Note that \eqref{infinite_SDP} contains an infinite number of LMI constraints and therefore cannot be solved numerically using a standard SDP solver. To address this we show how a conservative inner approximation of the feasible set can be defined in terms of a finite conjunction of LMIs. Before stating this result, we prove the following lemma which shows how an infinite family of LMIs can be conservatively approximated by a finite family of LMIs by assuming the existence of polytopic bounds on the coefficient matrices. 

\begin{lemma}\label{LMI_finite_approximation}
For all $i = 0, \dots, n$ let $F_i(z)$ be a family of symmetric matrices parameterized by $z \in Z$. Assume that there exist a finite set of matrices $\{F_{i k_i}:  k_i = 1, \dots, N_i; i=0, \dots, n\}$ such that for each $i$ the family $F_i(z)$ is bounded by the matrix polytope with vertices $F_{ik_i}$: 
\begin{equation}\label{coefficient_overapproximation}
 \{F_i(z) : z \in Z \} \subseteq Conv( \{ F_{i k_i} : k_i = 1, \dots, N_i\}) 
\end{equation}
where $Conv$ denotes the convex hull. Then 
\[
 F_{0 k_0} + x_1 F_{1 k_1} + \dots x_n F_{n k_n} \succeq 0 \quad \forall (k_0, \dots k_n) \in [N_0] \times \dots \times [N_n],
\]
where $[N_i]:=\{1,\dots,N_i\}$, implies
\[
F_0(z) + x_1 F_1(z) + \dots + x_n F_n(z) \succeq 0 \quad \forall z \in Z. 
\]
\end{lemma} 
\begin{proof}
Let $z \in Z$. Using the assumption \eqref{coefficient_overapproximation} we know that for each $i$ there exist nonnegative weights $\lambda_{ik_i}$ with $\sum_{k_i} \lambda_{ik_i} = 1$ such that $F_i(z) = \sum_{k_i} \lambda_{ik_i} F_{ik_i}$. Therefore 
\[
\begin{split}
F_0(z) & + x_1 F_1(z) + \dots + x_n F_n(z) \\
=& \sum_{k_0} \lambda_{0k_0} F_{0 k_0} + x_1 \sum_{k_1} \lambda_{1k_1} F_{1 k_1} + \dots 
    + x_n \sum_{k_n} \lambda_{nk_n} F_{n k_n} \\
=& \prod_{j \ne 0}\left( \sum_{k_j} \lambda_{jk_j} \right) \left(\sum_{k_0} \lambda_{0k_0} F_{0 k_0} \right) 
+ \dots + x_n  \prod_{j \ne n}\left( \sum_{k_j} \lambda_{jk_j} \right) \left(\sum_{k_n} \lambda_{nk_n} F_{n k_n}  \right) \\
=& \sum_{k_0} \dots \sum_{k_n} (\lambda_{0 k_0} \dots \lambda_{n k_n}) (F_{0 k_0} + x_1 F_{1 k_1} + \dots x_n F_{n k_n}) \\
\succeq & \ 0.  
\end{split}
\]
\end{proof}

We now state a result that allows us to find a set of weights satisfying \eqref{bound} by solving only a finite set of LMIs. 

\begin{proposition}\label{prop:finite_DSP} 
For each $i$ let $\{ E_{i \ell_i} : \ell_i = 1, \dots, n_i(n_i+1)/2\}$ be a basis for the space of $n_i \times n_i$ symmetric matrices. Suppose that there exist matrices such that
\[
\{ E_{i\ell_i} J_{ii}(t, x) + J_{ii}(t, x)^T E_{i \ell_i} : (t, x) \in [0, T] \times \mathcal{D} \} 
\subseteq 
Conv(\{F_{i \ell_i k_{i \ell_i}} : k_{i \ell_i} \in [N_{i \ell_i}] \}) 
\]
and 
\[
\{J_{ij}(t, x)^T E_{i \ell_i} J_{ij}(t, x) : (t, x) \in [0, T] \times \mathcal{D} \} 
\subseteq 
Conv(\{\tilde F_{ij \ell_i k_{ij \ell_i}} : k_{ij \ell_i} \in [N_{ij \ell_i}] \}).   
\]
Then any solution to the SDP
\begin{equation}\label{finite_SDP}
\begin{split}
 \textbf{find \quad} & x_{i \ell_i} \quad \forall i \in [k] \quad \forall \ell_i \in [n_i(n_i + 1)/2]  \\
 \textbf{subject to \quad} & \sum_{\ell_i} x_{i \ell_i} E_{i \ell_i} \succeq 0, 
 \quad \forall i \in [k] \\
 & \sum_{\ell_i} x_{i \ell_i} F_{i \ell_i k_{i \ell_i}} \preceq 2c_{ii} \sum_{\ell_i} x_{i \ell_i} E_{i \ell_i},  
 \quad \forall i \in [k] \\
 & \quad \quad \quad \quad \quad \quad \quad \quad \quad \quad \quad 
   \forall (k_{i1}, \dots, k_{i, n_i(n_i + 1)/2}) \in [N_{i1}] \times \dots \times [N_{i, n_i(n_i + 1)/2}] \\
 &  \sum_{\ell_i} x_{i \ell_i} \tilde F_{ij \ell_i k_{ij \ell_i}} \preceq c_{ij}^2 \sum_{\ell_j} x_{j \ell_j}  E_{jk_j} 
 \quad \forall i \in [k] \quad \forall j \in [k]\setminus\{i\}
 \\
 & \quad \quad \quad \quad \quad \quad \quad \quad \quad \quad \quad \forall (k_{ij1}, \dots, k_{ij, n_i(n_i + 1)/2}) \in [N_{ij1}] \times \dots \times [N_{ij, n_i(n_i + 1)/2}]
\end{split}
\end{equation}
yields a solution $\Gamma_i = \sum_{\ell_i} x_{i \ell_i} E_{i \ell_i}$ to \eqref{infinite_SDP}. 
\end{proposition}  

The proof follows in a straightforward manner from \eqref{infinite_SDP}  by expanding the decision variables as $\Gamma_i = \sum_{\ell_i} x_{i\ell_i} E_{i \ell_i}$ then applying Lemma \ref{LMI_finite_approximation}. Note that if $\mathcal{D}$ is compact and $(t, x) \mapsto J(t, x)$ is continuous, it is always possible to find a collection of such matrices $F$ and $\tilde F$. 

\section{Proof of Proposition \ref{main}}  \label{proof}
Let $\psi(t,s)$ denote the solution of (\ref{NL}) with initial condition $s x(0)+(1-s)z(0)$; that is,
\begin{eqnarray}\label{diffeq}
\frac{\partial \psi(t,s)}{\partial t}&=&f(t,\psi(t,s)) \\
\psi(0,s)&=&s x(0)+(1-s)z(0).\label{ic}
\end{eqnarray}
In particular,
\begin{equation}\label{ends}
\psi(t,1)=x(t) \quad \mbox{and} \quad \psi(t,0)=z(t).
\end{equation}
Taking the derivative of both sides of (\ref{diffeq}) with respect to $s$ we get
$$
\frac{\partial^2 \psi(t,s)}{\partial t\partial s}=\frac{\partial{f(t,\psi(t,s))}}{\partial s}=J(t,\psi(t,s))\frac{\partial \psi(t,s)}{\partial s},
$$
which means that the variable
\begin{equation}\label{wdef}
w(t,s)\triangleq \frac{\partial \psi(t,s)}{\partial s}
\end{equation}
satisfies
\begin{equation}\label{weq}
\frac{\partial w(t,s)}{\partial t} = J(t,\psi(t,s)) w(t,s).
\end{equation}
We then conclude from Lemma \ref{Coppel+} below that
\begin{equation}\label{adapt}
D^+|w_i(t,s)| \le \mu(J_{ii}(t,\psi(t,s))) |w_i(t,s)|+\sum_{j\neq i}\|J_{ij}(t,\psi(t,s))\| |w_j(t,s)|,
\end{equation}
where $D^+$ denotes the upper right-hand derivative with respect to $t$.
Since $\psi(t,s)\in \mathcal{D}$ for $t\in [0,T]$ and (\ref{bound}) holds for all $(t,x)\in [0,T]\times \mathcal{D}$, we conclude 
\begin{equation}\label{adapt2}
D^+|w_i(t,s)| \le C_{ii} |w_i(t,s)|+\sum_{j\neq i}C_{ij} |w_j(t,s)|.
\end{equation}
This means that 
\begin{equation}\label{combine}
D^+\left[ \begin{array}{c} |w_1(t,s)| \\ \vdots \\   |w_k(t,s)| \end{array}\right]\le C \left[ \begin{array}{c} |w_1(t,s)| \\ \vdots \\   |w_k(t,s)| \end{array}\right]
\end{equation}
 and, since the matrix $C$ is Metzler ($C_{ij}\ge 0$ when $i\neq j$),  it follows from standard comparison theorems for positive systems that
\begin{equation}\label{comp1}
\left[ \begin{array}{c} |w_1(t,s)| \\ \vdots \\   |w_k(t,s)| \end{array}\right]\le \exp(Ct) \left[ \begin{array}{c} |w_1(0,s)| \\ \vdots \\   |w_k(0,s)| \end{array}\right].
\end{equation}
Note from (\ref{wdef}) and (\ref{ic}) that
\begin{equation}
w(0,s)=\frac{\partial \psi(0,s)}{\partial s}=x(0)-z(0)
\end{equation}
and, thus,
\begin{equation}\label{interm1}
\left[ \begin{array}{c} |w_1(0,s)| \\ \vdots \\   |w_k(0,s)| \end{array}\right]=\left[\begin{array}{c} |x_1(0)-z_1(0)| \\ \vdots \\ |x_k(0)-z_k(0)| \end{array} \right].
\end{equation}
Substituting (\ref{interm1})  in (\ref{comp1}) we get
\begin{equation}\label{comp2}
\left[ \begin{array}{c} |w_1(t,s)| \\ \vdots \\   |w_k(t,s)| \end{array}\right]\le \exp(Ct) \left[\begin{array}{c} |x_1(0)-z_1(0)| \\ \vdots \\ |x_k(0)-z_k(0)| \end{array} \right].
\end{equation}
Next, note from (\ref{ends}) that
\begin{equation}
x(t)-z(t)=\psi(t,1)-\psi(t,0)=\int_0^1  \frac{\partial \psi(t,s)}{\partial s}ds=\int_0^1 w(t,s)ds,
\end{equation}
which implies
\begin{equation}\label{interm2}
\left[\begin{array}{c} |x_1(t)-z_1(t)| \\ \vdots \\ |x_k(t)-z_k(t)| \end{array} \right] \le \left[\begin{array}{c} \int_0^1|w_1(t,s)|ds \\ \vdots \\ \int_0^1 |w_k(t,s)|ds\end{array} \right].
\end{equation}
Noting from (\ref{comp2}) that
\begin{equation}\label{comp3}
\left[\begin{array}{c} \int_0^1|w_1(t,s)|ds \\ \vdots \\ \int_0^1 |w_k(t,s)|ds\end{array} \right]\le \exp(Ct) \left[\begin{array}{c} |x_1(0)-z_1(0)| \\ \vdots \\ |x_k(0)-z_k(0)| \end{array} \right]
\end{equation}
and combining with (\ref{interm2}) we obtain (\ref{contract}). \hfill $\Box$

\begin{lemma}\label{Coppel+}Consider the linear time-varying system
\begin{equation}\label{full}
\dot{w}(t)=A(t)w(t), \quad w(t)\in \mathbb{R}^n,
\end{equation}
where $A(\cdot)$ is continuous.  Suppose we decompose $A(t)\in \mathbb{R}^{n\times n}$  into blocks $A_{ij}(t)\in \mathbb{R}^{n_i\times n_j}$, $i,j=1,\dots,k$ such that  $n_1+\cdots+n_k=n$, and let $w_i(t)\in \mathbb{R}^{n_i}$, $i=1,\dots,k$, constitute a conformal partition of $w(t)\in \mathbb{R}^n$.
Then
\begin{equation}\label{comparison}
D^+|w_i(t)|\triangleq \lim_{h\rightarrow 0^+}\frac{|w_i(t+h)|-|w_i(t)|}{h} \le \mu(A_{ii}(t)) |w_i(t)|+\sum_{j\neq i}\|A_{ij}(t)\| |w_j(t)|.
\end{equation}
\end{lemma}

\noindent
{\it Proof of Lemma \ref{Coppel+}:}  Note that
\begin{eqnarray*}
 \lim_{h\rightarrow 0^+}\frac{|w_i(t+h)|-|w_i(t)|}{h} &=& \lim_{h\rightarrow 0^+}\frac{|w_i(t)+h\dot{w}_i(t)|-|w_i(t)|}{h} \\
 &=& \lim_{h\rightarrow 0^+}\frac{|w_i(t)+hA_{ii}(t){w}_i(t)+h\sum_{j\neq i}A_{ij}(t)w_j(t)|-|w_i(t)|}{h} \\
 &\le & \lim_{h\rightarrow 0^+}\frac{|w_i(t)+hA_{ii}(t){w}_i(t)|-|w_i(t)|}{h}+|\sum_{j\neq i}A_{ij}(t)w_j(t)| \\
  &\le & \lim_{h\rightarrow 0^+}\frac{\| I+hA_{ii}(t)\|-1}{h}|w_i(t)|+\sum_{j\neq i}\|A_{ij}(t)\|\,|w_j(t)|.
\end{eqnarray*}
Then (\ref{comparison}) follows from the definition of the matrix norm, (\ref{measure}). \hfill $\Box$

\end{document}